# Controlling Metal-Insulator Transitions in Vanadium Oxide Thin Films by Modifying Oxygen Stoichiometry

*Min-Han Lee[1,2,\*], Yoav Kalcheim[2], Javier del Valle[2,+], and Ivan K. Schuller[1,2]*

[1]*Materials Science and Engineering Program, University of California San Diego, La Jolla, California 92093, USA*

[2] *Department of Physics and Center for Advanced Nanoscience, University of California San Diego, La Jolla, California 92093, USA*

[+]*Present address: Department of Quantum Matter Physics, University of Geneva, Quai Ernest-Ansermet 24, 1205 Geneva, Switzerland*

**Abstract:**

**Vanadium oxides are strongly correlated materials which display metal-insulator transitions as well as various structural and magnetic properties that depend heavily on oxygen stoichiometry. Therefore, it is crucial to precisely control oxygen stoichiometry in these materials, especially in thin films. This work demonstrates a high-vacuum gas evolution technique which allows for the modification of oxygen concentration in $VO_X$ thin films by carefully tuning thermodynamic conditions. We were able to control the evolution between $VO_2$, $V_3O_5$, and $V_2O_3$ phases on sapphire substrates, overcoming the narrow phase stability of adjacent Magnéli phases. A variety of annealing routes were found to achieve the desired phases and eventually to control the metal-insulator transition (MIT). The pronounced MIT of the transformed films along with the detailed structural investigations based on x-ray diffraction measurements and reciprocal space mapping show that optimal stoichiometry is obtained and stabilized. Using this technique, we find that the thin film V-O phase diagram differs from that of the bulk material due to strain and finite size effects. Our study demonstrates new pathways to strategically tune the oxygen stoichiometry in complex oxides and provides a roadmap for understanding the phase stability of $VO_X$ thin films.**

# 1. INTRODUCTION

Metal-insulator transitions (MITs) in oxides have been of special interest in condensed matter physics[1-2] and technology[3] in past decades. Specifically, vanadium oxides (VO$_X$) are considered very promising materials for next-generation oxide electronics. For instance, VO$_2$ has attracted much attention owing to its MIT near room temperature (T$_{MIT}$ ~340 K), in which a large increase in electric resistivity is accompanied by a structural phase transition(SPT).[4-5] Additional interest in VO$_2$ concerns the possibility of modulating this transition by external stimuli.[6-8] The significant change in electrical resistivity offers a platform for resistive switching[9] and neuromorphic computation[10-12] related applications. In pure and Cr-doped V$_2$O$_3$, the concurrence of metal-insulator, structural and magnetic phase transitions have been the subject of much fundamental research.[13-16] The V$_3$O$_5$ phase, as one of the only two VO$_X$ members with a MIT above room temperature (T$_{MIT}$ ~430 K), is an excellent candidate for use in silicon-based technologies[17] and switching device applications.[18-20] During this MIT transition, the lattice structure remains monoclinic with a change in space group symmetry from I2/c to P2/c. The possibility to trigger the MIT without a significant structural change is a great advantage for V$_3$O$_5$ based switching devices.

Vanadium oxides are very sensitive to changes in oxygen stoichiometry. Vanadium is a multivalent element displaying oxidation states between V$^{2+}$ and V$^{5+}$, which gives rise to a series of compounds known as Magnéli (V$_n$O$_{2n-1}$) and Wadsley phases (V$_n$O$_{2n+1}$).[21-23] The existence of multiple phases leads to a complicated V-O phase diagram.[24-26] Furthermore, shrinking them into thin films or other nanoscale structures is needed to increase their functionality.[27] Typically, the V-O phase diagram is constructed with the thermodynamic data of bulk samples. However, as the size is reduced, the increase in the surface-to-volume ratio introduces an additional surface energy term in the Gibbs free energy that might affect the thermodynamically stable phase region.[28-29] Despite decades of research, a comprehensive understanding of thin film thermodynamics in the VO$_X$ system is still lacking. Most studies have only focused on the growth of VO$_2$ thin films and the improvement of their MIT properties by thermal annealing. [2,3,30-32] Other oxides such as V$_3$O$_5$, which is the closest Magnéli phase to V$_2$O$_3$, is particularly difficult to deposit directly because of the narrow range of allowable oxygen stoichiometry (with an O/V ratio between 1.666 and 1.668 ± 0.002).[33-34] The fabrication of V$_3$O$_5$ film in a recent work still contains small amount of V$_2$O$_3$ impurities.[35] In general, obtaining high quality thin films of a specific VO$_X$ and understanding the thin film phase diagram remains a challenging task.

In this study, we achieved control over oxygen stoichiometry in vanadium oxide thin films by using a high-vacuum gas evolution system. The experimental setup is shown in Fig. 1a. (see the experimental section for more details) We first sputtered VO$_2$ or V$_2$O$_3$ thin films on sapphire (α-Al2O3) substrates and then placed the samples in the center of a high vacuum tube furnace for heat treatment using ultra high purity oxygen. A similar technique has been used to control oxygen deficiency in high-temperature superconducting YBa$_2$Cu$_3$O$_{7-X}$ samples.[36-37] Starting from single-phase VO$_2$ or V$_2$O$_3$ films, we are able to transform the initial material into a different VO$_X$ film by carefully following a controlled route in the partial oxygen pressure (PO$_2$)-temperature (T) diagram. (Fig. 1b.) Various routes were tested to optimize the purity and enhance the MIT properties of the resulting compound. After treatment, the resulting phases were investigated using x-ray diffraction (XRD), reciprocal space mapping (RSM) and transport measurements. We show that our technique can provide excellent control over a wide-range of temperatures and oxygen partial pressures down to high vacuum conditions, allowing us to minimize the effect of other gases or contaminations. This offers significant advantages over normal thermal oxidation/reduction methods using gas mixture with limited vacuum range ($\geq$ 0.01 torr).[38] The precision of our method is shown by obtaining single-phase $V_3O_5$ thin films with the pronounced MIT despite their extremely narrow phase stability range. We are thus able to modulate the metal-insulator transitions between different phases and construct a detailed vanadium-oxygen phase diagram for thin $VO_X$ films.

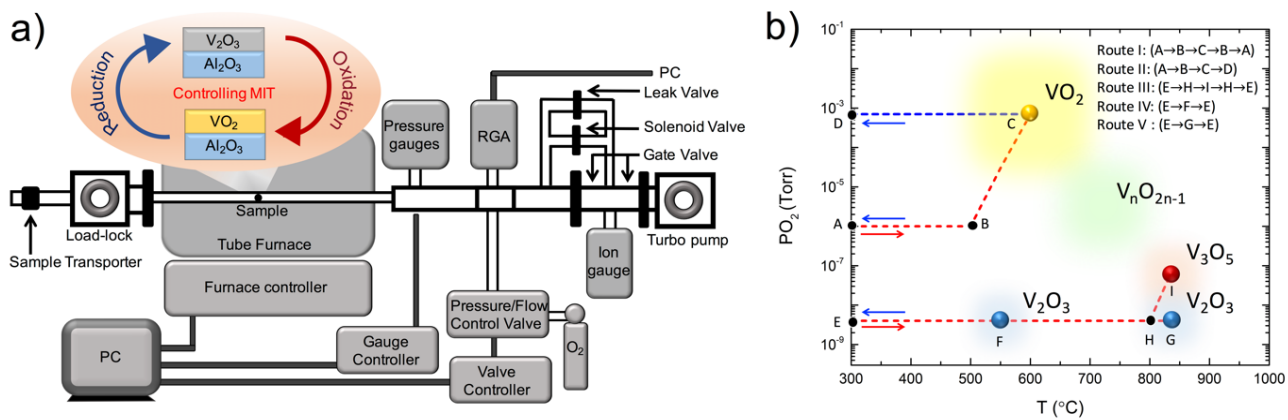

FIG. 1. *(a) Schematic illustration of the gas evolution setup used to control the oxygen stoichiometry in $VO_X$. (b) Synthesis pathways in the thermodynamic phase diagram of $VO_X$ thin films. The dashed lines represent the different heating/cooling routes for preparing various phases of vanadium oxide.*

## 2. RESULTS AND DISCUSSION

### 2.1 Evolution of $V_2O_3$ phase into $VO_2$

Figure 2a shows electrical transport of $V_2O_3$/$Al_2O_3$ (r-cut) thin films before and after different gas evolution treatments. The R-T curve of as-deposited $V_2O_3$ shows a sharp metal-insulator transition (MIT) at $T_{MIT}$ ~160 K, with at least a six orders of magnitude change in electrical resistance and a 10 K thermal hysteresis between the heating and cooling branches. The single-phase growth of $V_2O_3$ thin film on r-cut sapphire is confirmed by the XRD scan shown in Fig 2b. The scattering intensity of the $V_2O_3$ is much stronger than sapphire when we align the x-ray to the $V_2O_3$ peak. There are three out-of-plane diffraction peaks corresponding to (012), (024) and (036) corundum $V_2O_3$ phase. X-ray measurements on similar films [39] suggest that the $V_2O_3$ grows epitaxially along the crystallographic orientation of the (012) sapphire substrate orientation. By following the annealing process (route I) in the $PO_2$-T phase diagram (Fig. 1b), this film was transformed into $VO_2$, with a three orders of magnitude MIT at $T_{MIT}$ ~340K (Fig. 2a). Interestingly, this MIT was significantly improved and enhanced by following route II. The resulting $VO_2$ phase was confirmed by reciprocal space mapping (RSM) (see Fig. 2f and 2g). The absence of diffraction peaks from other Magnéli phases indicates the formation of single-phase VO$_2$ films.

Figure 2b shows the XRD spectra for (route I)-annealed and (route II)-annealed samples. XRD patterns for samples subjected to both routes, clearly show the diffraction peak at 37.1°, corresponding to the (200) plane of the VO$_2$ monoclinic structure. The resulting VO$_2$ films still exhibit preferred orientation with the (200) plane pointing at ~40° to the surface normal ($\chi$~ 40º), while for the directly sputtered VO$_2$ films on r-cut sapphire, the (200) plane is parallel to the substrate surface ($\chi$~ 0º) (Fig. 4b). The formation of a tilted (200) plane indicate the lack of well-aligned epitaxial layers and the appearance of disorder at the VO$_2$/Al$_2$O$_3$ interface, as present in many oxide heterostructures.[40] It should be noted that there are structural similarities between rutile VO$_2$ and corundum V$_2$O$_3$. Previous studies have shown that through a common parent structure with hexagonal-close-packing arrangement, VO$_2$ and V$_2$O$_3$ may be derived from each other through a well-defined symmetry-breaking mechanism along the [001] direction of corundum V$_2$O$_3$ and the [100] of rutile VO$_2$. [41] It is thus unlikely that formation of VO$_2$ from V$_2$O$_3$ would be randomly oriented.

The AFM images demonstrate differences in surface morphology and roughness between the as-deposited V$_2$O$_3$ (Fig. 2c), (route I)-annealed film (Fig. 2d) and (route II)-annealed film (Fig. 2e). The root-mean-square (RMS) roughness extracted from the Fig. 2c, 2d and 2e was 5.62 nm, 6.22 nm and 8.29 nm, respectively. The original V$_2$O$_3$ film prepared by sputtering reveals a relatively smooth surface. The quenching process along route I results in a bigger grain size (~190 nm assuming spherical shape) of the VO$_2$ film. It is this (route I)-annealed VO$_2$ film which exhibits a smaller width in the hysteresis loop appearing in the R-T curve at the MIT (Fig. 2a). Consequently, our results suggest that the hysteresis loop at the MIT broadens with decreasing the grain size.[42-43] Earlier reports show that smaller grains lead to a high density of grain boundaries resulting in a decrease in the magnitude of the resistivity change across the MIT.[44] Alternatively, the larger resistance switching ratios and smaller grains (~75 nm) observed in (route II)-annealed VO$_2$ film, suggest that the change in electrical resistance cannot be solely due to a grain size effect. Oxygen stoichiometry must also play a role in the MIT behavior of VO$_2$. The activation energy (E$_a$) of VO$_2$ insulating state (T ≤ 310 K) can be estimated from the linear fitting of the ratio ln R(T)/(1/k$_b$T), as R(T) = R$_o$ exp (E$_a$/k$_b$T). The results obtained from (route I)-annealed and (route II)-annealed VO$_2$ are ~191 meV and ~259 meV, respectively. Quenching the (route I)-annealed VO$_2$ film under relatively low PO$_2$ is likely to increase the density of defects and promote oxygen loss. These vacancies might have the effect of doping in the VO$_2$, thereby increasing the number of free charge carriers and reducing the magnitude of the resistive change across the MIT.[45,46] Detailed phase stability diagram (Fig. 6.) also shows that annealing the sample in point C at 600 ℃ with low oxygen pressure will result in oxygen–deficient VO$_{2-x}$ phases.

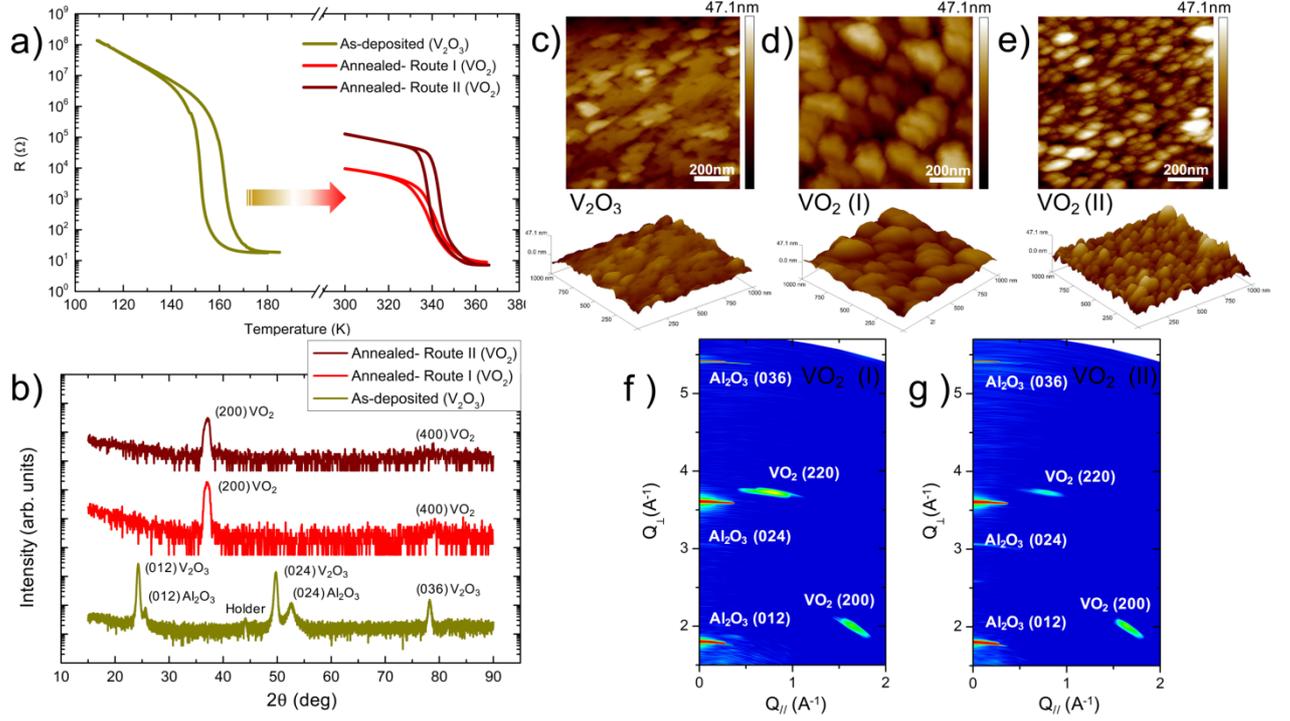

FIG. 2. *Formation of $VO_2$ phase from $V_2O_3$ thin film on (012) r-cut sapphire substrate. (a) Resistance vs. temperature of vanadium oxide thin films under various oxidation conditions. After annealing, the thin films showed a sharp $VO_2$ metal-insulator transition at ~340 K. (b) Room-temperature XRD scans for vanadium oxide films. The different diffraction spectra correspond to the $V_2O_3$ phase (as-deposited) and the $VO_2$ phase (annealed following route I or route II). The crystallographic orientation of the $VO_2$ film was (200), which points along ~40° to the surface normal ($\chi$ ~ 40°). Angle ($\chi$) denotes the degree between surface normal and the diffraction plane. AFM images (1 × 1 µm$^2$) showing the surface morphology of (c) as-deposited film, (d) (route I)-annealed film and (e) (route II)-annealed film. The scale bar is 200 nm for all images. Reciprocal space mapping (RSM) of (f) (route I)-annealed and g) (route II)-annealed $VO_2$ films. $Q\perp$ denotes the out-of-plane orientation (012) of $Al_2O_3$.*

## 2.2 Evolution of $V_2O_3$ phase into $V_3O_5$

Due to the numerous possible oxidation states, vanadium oxides are very sensitive to changes in oxygen content. Here, we show that oxidation of $V_2O_3$ to form the $V_3O_5$ can be achieved on c-cut and r-cut sapphire by slightly increasing the $PO_2$ at ~840 °C. Figure 3a displays the temperature dependent electrical transport of a $V_2O_3$ film grown on c-cut sapphire before and after the gas evolution treatment. The R-T curve of the as-deposited $V_2O_3$ film shows a suppressed metal-insulator transition with only a slight resistance variation. We note that this $V_2O_3$ sample was grown under the same conditions as the one shown in Fig. 2a, which shows a much more pronounced MIT. This suppression of the MIT in $V_2O_3$ grown on c-cut sapphire had been previously attributed to strain and microstructural effects between thin film and substrate.[47,48] The lattice parameters and the c/a ratio of our epitaxial c-cut $V_2O_3$ thin film are ~2.81 (a=b= 4.98 Å; c=14.01 Å). This ratio is consistent with the reported value of c-cut $V_2O_3$ with unusual metal-insulator-metal behavior.[47] In previous work, we have shown that the structural phase transition and metal-insulator transition were robustly coupled to each other.[16] Therefore, it is likely that the structural phase transition in our $V_2O_3$ sample is highly dependent on the a-and c-axis deformation as well. And this leads to the suppression of the MIT. The XRD pattern in Fig. 3b shows that corundum $V_2O_3$ film is highly oriented with the (001) c-cut sapphire substrate. By annealing this film along route III (Fig. 1b), we were able to transform it into a single-phase $V_3O_5$ film with an MIT of more than one order of magnitude at $T_{MIT}$ ~417 K. The smooth MIT behavior above room temperature together with the absence of thermal hysteresis in the R(T) are signs of a second-order phase transition, as reported previously.[18,49]

The presence of a single-phase polycrystalline structure is confirmed by RSM as shown in Fig. 3e. The axis $Q_\perp$ points along the out-of-plane orientation (001) for the $Al_2O_3$ substrate. Figure 3b shows six diffraction peaks at 2θ = 36.0°, 38.5°, 38.9°, 76.2°, 82.3° and 83.4° corresponding to the (020), (202), (400), (040), (404), and (800) planes in $V_3O_5$, respectively.[50] The AFM image of the as-deposited $V_2O_3$ film (Fig. 3c) indicates a relatively smooth film surface with 0.72 nm RMS roughness. The resulting $V_3O_5$ film shown in Fig. 3d has a much rougher surface with 13.1 nm RMS roughness. Furthermore, the grain structure is highly irregular after the treatment, as suggested by the XRD and RSM data.

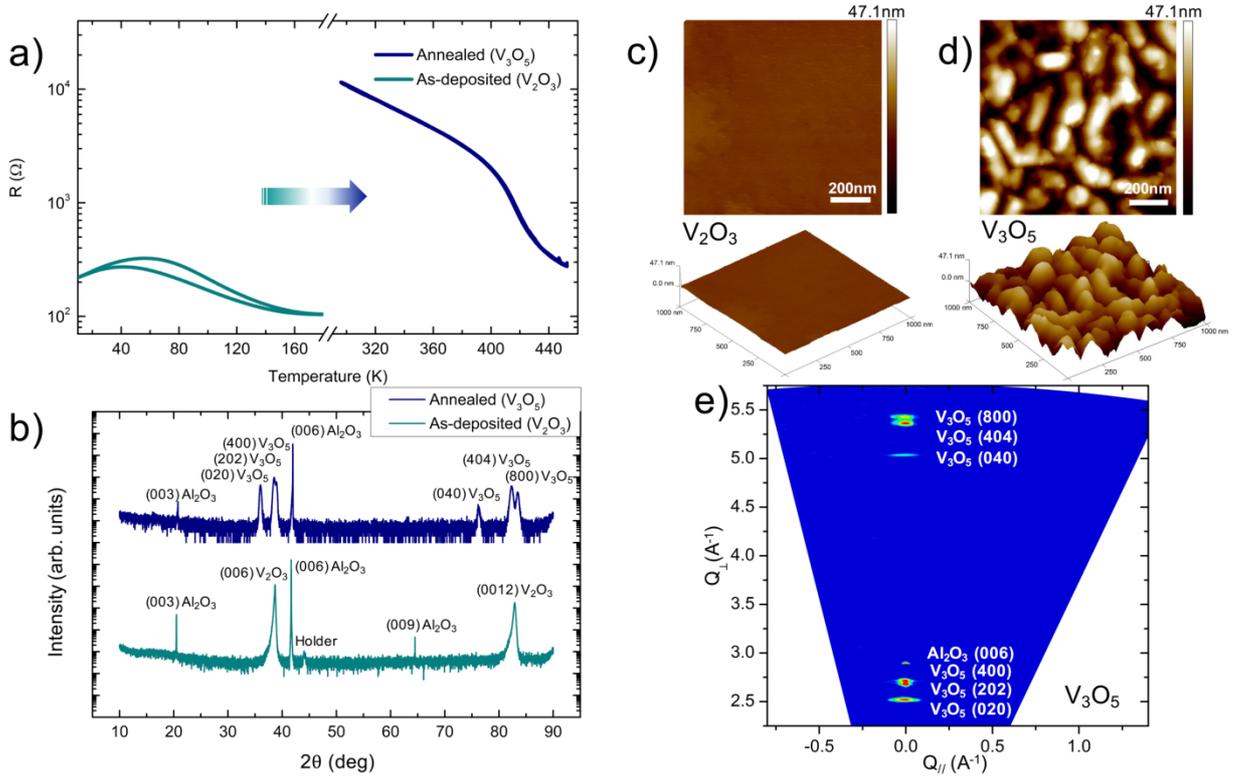

FIG. 3. *Formation of $V_3O_5$ phase from $V_2O_3$ thin film on (001) c-cut sapphire substrate. (a) Resistance as a function of temperature of $V_2O_3$ film and resulting $V_3O_5$ film. The MIT temperature ($T_{MIT}$) of the $V_3O_5$ is ~417 K. (b) Room-temperature XRD scans. Evolution of different peaks corresponding to $V_2O_3$ phase (as-deposited) and $V_3O_5$ phase (annealed following route III). Preferred crystallographic orientations of $V_3O_5$ film were (020), (202) and (400). AFM images revealed the surface morphology of (c) as-deposited film and (d)*

*(route III)-annealed film. (e) X-ray RSM of the resulting $V_3O_5$ film. $Q\perp$ denotes the out-of-plane orientation (001) of $Al_2O_3$ (perpendicular to the film surface).*

Figure 4a demonstrates the electrical transport of a $V_2O_3$ film on r-cut sapphire before and after gas evolution treatments along route III. After annealing at point I for 3 hours we observe an increase in resistance in the metallic states and a reduction of the magnitude of the hysteresis. However, the large MIT at $T_{MIT}$~150 K indicates that the major phase in the film is still $V_2O_3$. The formation of pure $V_3O_5$ phase is achieved by a second annealing process for an additional 6 hours along route III. The R(T) starts showing a smooth MIT above room temperature ($T_{MIT}$~404 K) without thermal hysteresis. Comparison of R-T curves of $V_3O_5$ films grown on c-cut (001) and r-cut (012) $Al_2O_3$ substrates is shown in Fig. 4b. Both samples reveal an MIT of more than one order of magnitude. For ease of comparison, the results were normalized by the metallic state resistance ($R_{450K}$). Since the transformed $V_3O_5$ films were polycrystalline, it is likely that the strain in the different grains is highly inhomogeneous. Different sapphire substrates will generate different dislocations and strain. Due to the numerous grain orientations, it is quite challenging to precisely quantify the strain in the polycrystalline thin film. By comparing the lattice spacings in $V_3O_5$ grown on r-cut and c-cut oriented sapphire, the structural distortion can still be estimated. Compared to the c-cut $V_3O_5$ thin film, the relative strains for the (100), (010), (001) planes of $V_3O_5$/r-cut sapphire are considerably more compressed ($\Delta d_{100} \sim -0.71\%$; $\Delta d_{010}\sim 0.03\%$; $\Delta d_{001}\sim -2.1\%$). It is likely that larger compressive strain leads to lower $T_{MIT}$ as found in a previous study which has shown that $T_{MIT}$ in $V_3O_5$ decreases with pressure.[51] The strain generated from the sapphire substrates and the difference in microstructure and grains might lead to the 14 K variation in $T_{MIT}$. Figure 4c shows the evolution of the crystalline structure for the $V_2O_3$ film. After 3 hours annealing, the intensity of the diffraction peaks decreased, but the thin film still remained corundum $V_2O_3$. On the other hand, the second, lengthier annealing time (6 hours) allowed for a complete transformation into $V_3O_5$. Besides the out-of-plane sapphire peaks, one diffraction peak from the $V_3O_5$ (310) plane was found at $\chi \sim 17°$ (Fig. 4c and 4e). And several in-plane $V_3O_5$ peaks can be found using reciprocal space mapping. The RMS roughness of the transformed $V_3O_5$ film is 5.8 nm (Fig. 4d), close to the value of original sputtered $V_2O_3$ on r-cut sapphire (RMS roughness~5.62 nm).

Interestingly, a large difference in the annealing times required to form $V_3O_5$ is observed for the two different substrates (012) and (001) $Al_2O_3$. This indicates the crystallographic orientation of the film significantly affects thermodynamic kinetics and oxygen diffusion processes. As we observe from our x-ray diffraction measurements (Fig. 3b and 4c), for both substrates, in the final stage the samples only showed diffraction peaks associated with $V_3O_5$ (aside from the sapphire peaks) indicating high phase purity.

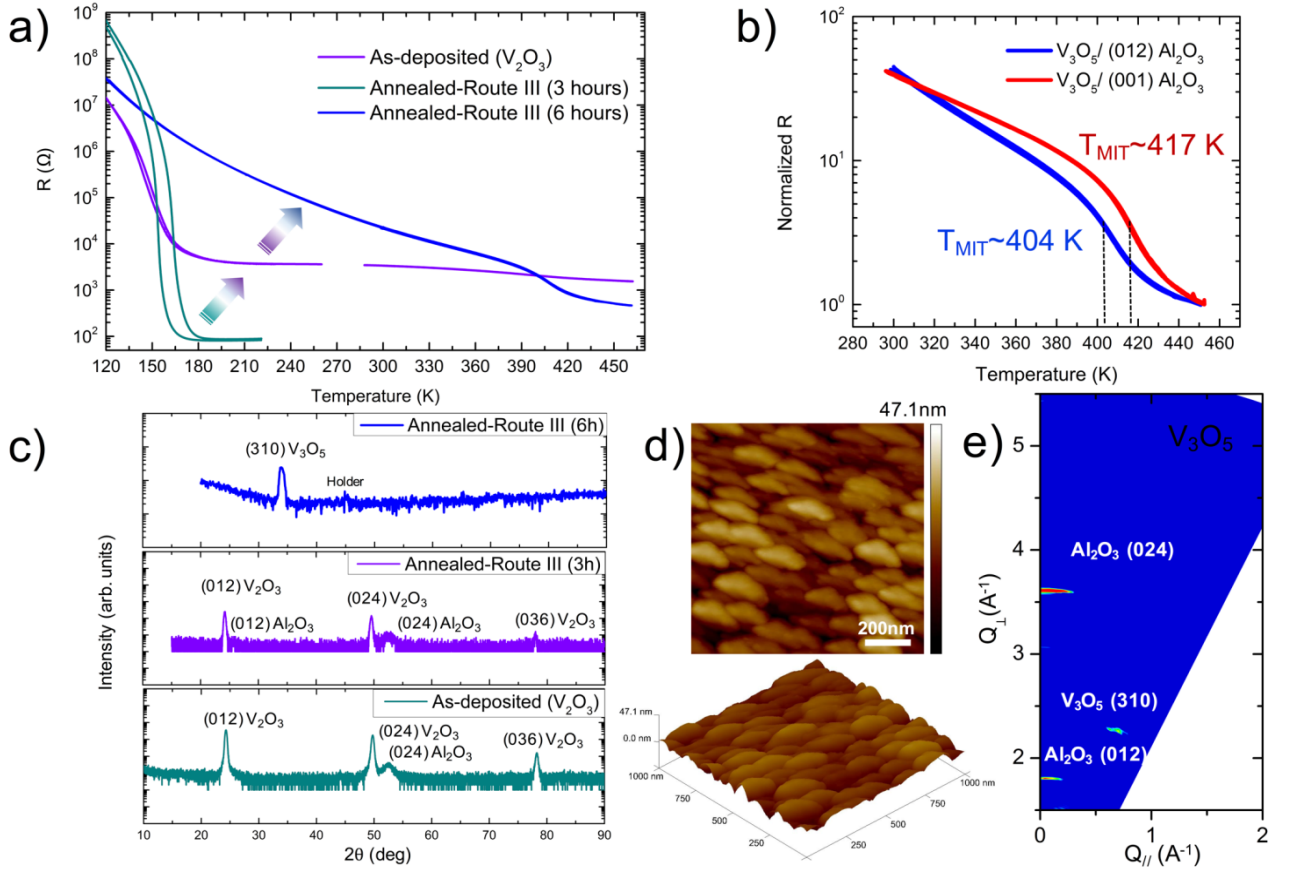

FIG. 4. *Formation of $V_3O_5$ Phase from $V_2O_3$ thin film on (012) r-cut sapphire substrate. (a) Resistance as a function of temperature for a $V_2O_3$ film: as deposited film (cyan); after annealing for 3 hours at point I (purple); after annealing for additional 6 hours at point I (blue). (b) Comparison of the normalized R(T) for $V_3O_5$ films on r-cut sapphire (blue) and c-cut sapphire (red). The black dashed lines mark the $T_{MIT}$ for the two different samples. (c) XRD measurements of samples after time annealed $V_2O_3$ thin film following route III. Preferred crystallographic orientation of the resulting $V_3O_5$ film was (310), which point along ~17° to the surface normal ($\chi$ ~ 17°). (d) AFM images revealed the surface morphology (e) X-ray RSM of the resulting $V_3O_5$ film. $Q\perp$ denotes the out-of-plane orientation (012) of $Al_2O_3$ (perpendicular to the film surface).*

## 2.3 Evolution of $VO_2$ phase into $V_2O_3$

In addition to the controlled oxidation of as-grown $V_2O_3$ thin films, we have also demonstrated and studied the controlled reduction of as-grown $VO_2$ thin films on r-cut sapphire. Figure 5a shows the electrical transport of a $VO_2$ film on an r-cut sapphire both before and after the gas evolution treatments along route IV. The R-T curve of as-deposited $VO_2$ shows a sharp metal-insulator transition (MIT) at $T_{MIT}$ ~340 K, with more than a three orders of magnitude resistance change. The single-phase growth of $VO_2$ is confirmed by the XRD scan shown in Fig. 5b. Two reflection peaks corresponding to the (200) and (400) planes for monoclinic $VO_2$ are observed at 2θ = 37.1° and 79.0°, respectively. By following the annealing process (route IV) shown in Fig. 1b, the highly textured $VO_2$ film was transformed into single-phase $V_2O_3$, now with a five orders of magnitude MIT with $T_{MIT}$~145 K. (see Fig. 5a) This value of the transition temperature $T_{MIT}$ ~145K, is slightly lower than the value of the transition temperature $T_{MIT}$ ~160K, for the directly sputtered $V_2O_3$ film, indicating that there might be a comparatively higher oxygen concentration in the transformed $V_2O_3$ film.[52] Moreover, we find no significant differences ($\Delta 2\theta \leq 0.035°$) in the out-of-plane (012) and in-plane (104) XRD peaks between directly sputtered $V_2O_3$ and transformed $V_2O_3$ films, ruling out strain effects as a major cause for the decrease in $T_{MIT}$. Previous studies on $V_2O_{3+x}$ have shown that a 15K variation in $T_{MIT}$ corresponds to an excess of oxygen concentration (x) of less than with 0.013. [53,54] This provides an estimate of the accuracy of our control over oxygen content in a gas evolution process. The presence of an epitaxial $V_2O_3$ film was confirmed by XRD (Fig. 5b) and RSM scans (Fig. 5e). Both in-plane and out-of-plane crystallographic orientations in $V_2O_3$ follow the substrate orientations. This indicates that the structural similarity between corundum $V_2O_3$ and $Al_2O_3$ promotes epitaxial growth during the atomic rearrangement. Therefore, it is reasonable to conclude that the coupling to the $Al_2O_3$ substrate plays an important role in determining the crystallographic orientations of the transformed film. We note that this is in contrast to the case of $VO_2$ derived from a $V_2O_3$ film, which develops a different orientation after gas evolution compared to the as-grown $VO_2$ film. This may be due to the higher degree of compatibility between the $Al_2O_3$ substrate and the isostructural $V_2O_3$ compared to the rutile $VO_2$.

For the transformation of as-grown $VO_2$ films into single-phase $V_2O_3$, it takes 12 hours (point F in the $PO_2$-T diagram shown in Fig. 1b) to reach an equilibrium state. We found that the as-grown $VO_2$ films that were subjected to a shortened annealing process of only three or six hours along route IV did not exhibit any clear MIT behavior and no XRD peaks can be found. This suggests that thermodynamic equilibrium may not have been reached in this shorter period of time. However, the reverse transformation of as-grown $V_2O_3$ films into single-phase $VO_2$ films requires a shorter annealing time of three hours. The discrepancy between these two cases suggests that the phase evolution mechanisms are also associated with kinetics and diffusion processes which are beyond the scope of this work. Furthermore, the monoclinic $VO_2$ film could also be transformed into corundum $V_2O_3$ phase by following an alternate annealing treatment along route V. This indicates that point G in Fig. 1b is still in the $V_2O_3$ stable region. Figures 5c and 5d show the surface morphology of the as-deposited film and (route IV)-annealed film, respectively. The $VO_2$ film prepared by sputtering shows a textured surface with the RMS roughness equal to 5.98 nm. The resulting $V_2O_3$ after the loss of oxygen shows a relatively rougher surface with roughness up to 8.35 nm.

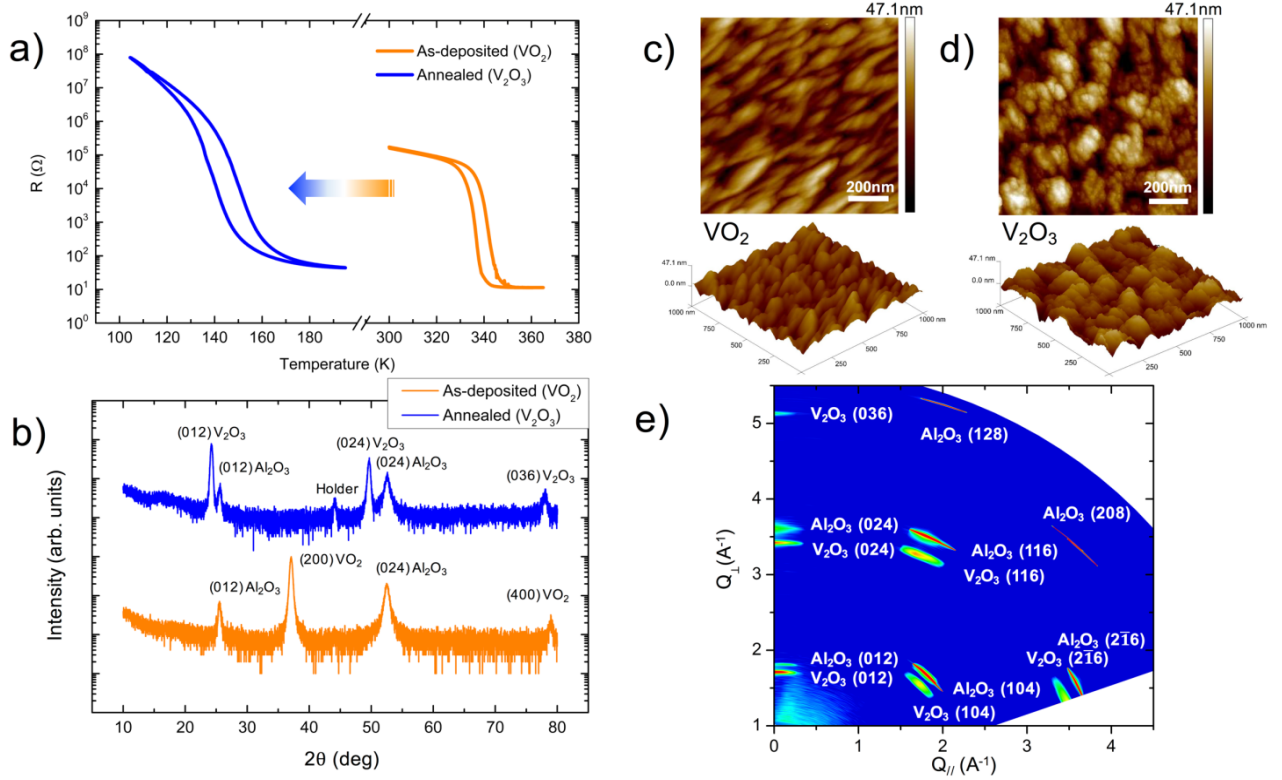

FIG. 5. *Formation of $V_2O_3$ Phase from $VO_2$ thin film on (012) r-cut sapphire substrate. (a) Resistance as a function of temperature of $VO_2$ film and resulting $V_2O_3$ film. (b) Room-temperature XRD scans. Evolution of different peaks corresponding to $VO_2$ phase (as-deposited) and $V_2O_3$ phase (following annealing route IV). AFM images revealed the surface morphology of (c) as-deposited film and (d) annealed film. (e) Wide-range RSM of the transformed $V_2O_3$ film. Both in-plane and out-of-plane crystallographic orientations of the $V_2O_3$ film strongly followed the r-cut sapphire substrate orientations, suggesting the formation of epitaxial structures after gas evolution. $Q\perp$ denotes the out-of-plane orientation (012) of $Al_2O_3$.*

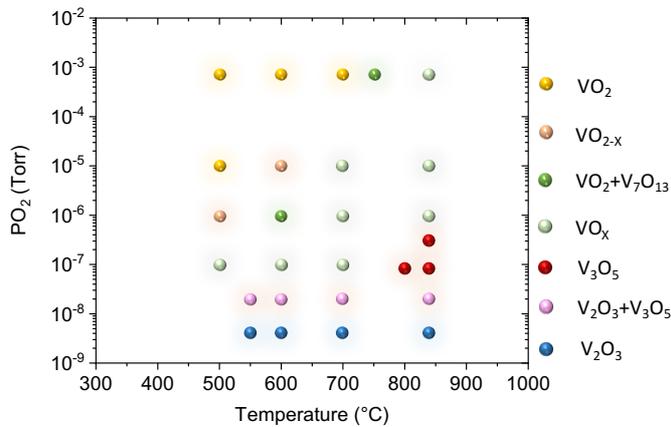

FIG. 6. *V-O phase diagram for thin film. Circles represent the tested annealing condition.*

Our technique can provide excellent control over a wide-range of temperatures and oxygen partial pressures, allowing us to explore a more complete picture of thin film phase diagram. It is worth noting that the V-O phase diagram (Fig. 6) we obtain starting from 100 nm $V_2O_3$ is substantially different from the bulk V-O phase diagram.[24-26] Among all the Magnéli phases, the only thin film we could obtain in pure form was $V_3O_5$ and the annealing parameter range was very narrow. Moreover, starting from from 10 nm thin films, despite tens of trials, we could not obtain $V_3O_5$ under the same annealing conditions. These condition differences as compared to the bulk phase diagram may be due to variations in nanoscale surface energy and the lattice interaction with the substrate. This may also explain differences in required annealing time found for sapphire substrates of different orientations (c-cut and r-cut) as described above.

## 3. CONCLUSION

In summary, we have developed a high-vacuum gas evolution technique that allows for the precise modification of oxygen stoichiometry in $VO_X$ thin films. Using temperature-dependent electrical transport measurements, x-ray diffraction and reciprocal space mapping, we show that optimal oxygen stoichiometry is obtained in each of the transformed films. Based on our technique, the thermodynamic phase diagram can be well-defined and elucidated for the V-O system when scaling down the $VO_X$ from bulk into thin film form. Our results may play an important role in improving and controlling transport properties across the MIT in oxide-based devices. It may open a new way to synthesize exotic or Magnéli phase oxide films which cannot be directly grown by standard deposition techniques. We hope our work will encourage the use of gas evolution methods in other complex oxide systems.

## ACKNOWLEDGMENTS

The authors acknowledge support from the Air Force Office of Scientific Research grant FA9550-14-1-0202, supported by the Vannevar Bush Faculty Fellowship program, sponsored by the Basic Research Office of the Assistant Secretary of Defense for Research and Engineering and funded by the Office of Naval Research through grant N00014-15-1-2848. J.d.V. thanks Fundación Ramón Areces for their funding.

# 4. EXPERIMENTAL

## *Thin-film growth*

Single-phase $VO_2$ and $V_2O_3$ thin films were deposited on 7×12 mm$^2$ (012) r-cut or (001) c-cut sapphire (α-$Al_2O_3$) substrates using RF magnetron sputtering from a $V_2O_3$ target. As described in previous studies[9,39], the growth of $VO_2$ was done at a substrate temperature of 520 ℃ in an environment of Ar/$O_2$ mixture (8% $O_2$) at 3.7 mTorr. The $V_2O_3$ thin films were prepared at a 700 ℃ substrate temperature in an environment of ultrahigh purity Ar (>99.999%) at 8 mTorr.

## *Gas evolution system*

A high-vacuum oxygen annealing method was developed to synthesize complex oxide thin films with controlled oxygen stoichiometry. The experimental setup is shown in Fig. 1a. The base pressure in the system is ~1×10$^{-7}$ torr. Thin film samples were placed into the center of a Lindberg/Blue M$^{TM}$ 1200℃ tube furnace for heat treatment under different oxygen atmospheres. The $PO_2$ was measured using a Residual Gas Analyzer equipped with a quadrupole mass spectrometer (UHV to 10$^{-4}$ torr) and two capacitance manometers (10$^{-4}$ torr to 1000 torr). A MKS Series 245 pressure/flow control valve, a variable leak valve and a solenoid valve were used to carefully control the $PO_2$. High purity oxygen (>99.99%) was introduced into the annealing chamber using the computer-controlled metal-seated valve. This gas evolution setup provides a wide pressure range (UHV to 1000 torr) and allows for the continuous variation of the $PO_2$ under high vacuum conditions.

## *Evolution of $V_2O_3$ phase into $VO_2$*

Oxidation of as-deposited $V_2O_3$/$Al_2O_3$ (r-cut) thin films with 75 nm thickness was performed in the gas evolution system by continuously controlling $PO_2$ and T. The annealing conditions were chosen according to the known bulk vanadium-oxygen phase diagram.[24-26] By carefully following thermodynamically stable paths, different oxidation states of vanadium could be obtained from $V_2O_3$. At the start of route I (A→B→C→B→A) (see Fig. 1b), $PO_2$ was fixed to 1 ×10$^{-6}$ torr and the temperature was ramped up to 500 ℃ at a heating rate of 10 ℃/min. After reaching 500 ℃, the $PO_2$ was continuously increased along (B→C) in order to keep the sample in the potentially stable region of the $VO_2$ phase in $PO_2$-T diagram. The sample was then annealed at 600 ℃ for three hours with $PO_2$ ~7 × 10$^{-4}$ torr (at point C in Fig. 1b). After reaching thermodynamic equilibrium, both $PO_2$ and T were reduced at a cooling rate of 6 ℃/min along the return path (C→B). After reaching 500 ℃, the sample was then quenched to room temperature by transferring it to the load lock.

Similarly, by following route II (A→B→C→D), the oxygen stoichiometry in as-deposited $V_2O_3$ can be modified, leading to a transformation into a higher oxidation state. After a three-hour reaction at 600 ℃ (point C), the sample was cooled down to room temperature under continuous oxygen flow at $PO_2$ ~7 × 10$^{-4}$ torr. As will be discussed later, this method produces high-quality VO$_2$ thin films with sharp metal-insulator transitions. The ability to continuously and gradually control both PO$_2$ and T is important for maintaining thermodynamic equilibrium in the thin film sample at all times so as to prevent the formation of possible secondary phases during the process.

## *Evolution of V$_2$O$_3$ phase into V$_3$O$_5$*

For the oxidation reaction of V$_2$O$_3$ into V$_3$O$_5$, we used a similar procedure. At the start of route III (E→H→I), an as-deposited V$_2$O$_3$/Al$_2$O$_3$ (c-cut) film with thickness of approximately 100 nm was mounted in the tube furnace and the PO$_2$ was fixed to ~4 × 10$^{-9}$ torr. The temperature was then increased from RT to 800 ℃ at a rate of 10 ℃/min (E→H). From point H, both the PO$_2$ and temperature (T) were continuously increased along the possible stable region for V$_3$O$_5$. The thin film was then annealed for three hours at 838 ℃ with PO$_2$ ~8 ×10$^{-8}$ torr (at point I). After reaching phase equilibrium, both PO$_2$ and T were continuously reduced along the return route (I→H) at a cooling rate of 6 ℃/min and the sample was quenched from 800 ℃ to room temperature along the return path (H→E). For the case of V$_2$O$_3$/Al$_2$O$_3$ (r-cut) film, we performed the same experiment: the 100 nm film was annealed following route III. The annealing time was 3 hours at point I in the first round and increased to 6 hours for the second round.

## *Evolution of VO$_2$ phase into V$_2$O$_3$*

In order to understand the phase stability of VO$_2$, as-deposited VO$_2$/Al$_2$O$_3$ (r-cut) films with thickness of 75 nm were heat-treated under two different thermodynamic paths: route IV (E→F→E) and route V (E→G→E). At the start of route IV, the temperature was increased at a rate of 10 °C/min to 550 °C under PO$_2$ ~4 × 10$^{-9}$ torr. The annealing time was optimized to 12 hours. After the reduction reaction, the sample was then rapidly quenched in order to prevent the formation of other Magnéli phases that might occur during a slow cooling process. Similarly, along thermodynamic route V, the sample was stabilized at 838 °C for 12 hr (point G) and also rapidly quenched to room temperature.

## *Characterizations*

Room-temperature θ-2θ x-ray diffraction (XRD) and reciprocal space mapping (RSM) were used to characterize the lattice structure and its relation to the substrate structure using a Rigaku SmartLab x-ray diffractometer with Cu Kα radiation (λ = 1.54 Å). Atomic force microscopy (AFM) for room-temperature surface characterization was performed using a Veeco Scanning Probe Microscope (SPM). The scan area of each of the films was fixed to 1×1 μm$^2$. Temperature-dependent electrical transport measurements using a standard four-point probe configuration were performed in a Lakeshore TTPX probe station with a Keithley 6221 current source and a Keithley 2182A nanovoltmeter.